\documentclass[reprint,superscriptaddress,amsmath,amssymb,aps,pre]{revtex4-1}

\usepackage{graphicx}
\usepackage{dcolumn}
\usepackage{bm}
\usepackage{color}
\usepackage{bbold}
\usepackage{soul}
\usepackage{epsfig,epic}
\newcommand{\newc}{\newcommand}
\newc{\beq}{\begin{equation}}
\newc{\eeq}{\end{equation}}
\newc{\kt}{\rangle}
\newc{\br}{\langle}
\newc{\beqa}{\begin{eqnarray}}
\newc{\eeqa}{\end{eqnarray}}
\newc{\longra}{\longrightarrow}

\makeatletter
\let\Hy@backout\@gobble
\makeatother

\begin{document}

\title{Entanglement and localization transitions in eigenstates of interacting chaotic systems}

\author{Arul Lakshminarayan}
\affiliation{Max-Planck-Institut f\"ur Physik komplexer Systeme, N\"othnitzer Stra\ss{}e 38, 01187 Dresden, Germany}
\affiliation{Department of Physics, Indian Institute of Technology Madras, Chennai, India~600036}
\author{Shashi C. L. Srivastava}
\affiliation{Max-Planck-Institut f\"ur Physik komplexer Systeme, N\"othnitzer
Stra\ss{}e 38, 01187 Dresden, Germany}
\affiliation{Variable Energy Cyclotron Centre, 1/AF Bidhannagar, Kolkata 700064, India.}
\author{Roland Ketzmerick}
\affiliation{Max-Planck-Institut f\"ur Physik komplexer Systeme, N\"othnitzer
Stra\ss{}e 38, 01187 Dresden, Germany}
\affiliation{Technische Universit\"at Dresden, Institut f\"ur Theoretische
             Physik and Center for Dynamics, 01062 Dresden, Germany}

\author{Arnd B\"acker}
\affiliation{Max-Planck-Institut f\"ur Physik komplexer Systeme, N\"othnitzer
Stra\ss{}e 38, 01187 Dresden, Germany}
\affiliation{Technische Universit\"at Dresden, Institut f\"ur Theoretische
             Physik and Center for Dynamics, 01062 Dresden, Germany}
\author{Steven Tomsovic}
\affiliation{Max-Planck-Institut f\"ur Physik komplexer Systeme, N\"othnitzer
Stra\ss{}e 38, 01187 Dresden, Germany}
\affiliation{Technische Universit\"at Dresden, Institut f\"ur Theoretische
             Physik and Center for Dynamics, 01062 Dresden, Germany}
\affiliation{Department of Physics and Astronomy, Washington State University, Pullman, WA~99164-2814}

\date{\today}

\begin{abstract}
The entanglement and localization in eigenstates of strongly chaotic subsystems are studied as a function of their interaction strength.  Excellent measures for this purpose are the von-Neumann entropy, Havrda-Charv{\' a}t-Tsallis  entropies, and the averaged inverse participation ratio.  All the entropies are shown to follow a remarkably simple exponential form, which describes a universal and rapid transition to nearly maximal entanglement for increasing interaction strength.  An unexpectedly exact relationship between the subsystem averaged inverse participation ratio and purity is derived that infers the transition in the localization as well.
\end{abstract}

\pacs{PACS here}

\maketitle

Entanglement is a central non-classical feature of quantum mechanics.  It has been the subject of a broad range of studies from quantum information protocols such as teleportation~\cite{Bennett93} to quantum phase transitions~\cite{Amico08} and, curiously, system-environment entanglement gives rise to emergent classical behavior~\cite{Zurek91,Zurek03}. Very useful entanglement measures are provided by the von-Neumann and Havrda-Charv\'at-Tsallis entropies~\cite{Bennett96,Havrda67,Tsallis88,Bengtsson}.  A critically important aspect is that its production is often necessary for quantum computing~\cite{NielsenChuang10}.   Nevertheless, entanglement in eigenstates of interacting strongly chaotic subsystems is not well understood.  A fundamentally interesting question is thus, `'how entangled are strongly chaotic particles as a function of their interaction strength?''.

Work involving entropy production has a long history.  Simple models have been studied to find the evolution of entanglement in initially separable states, e.g.~a two-state system coupled to a many-state random Hamiltonian~\cite{Albrecht92}. Entanglement production in coupled systems, whose classical limits display integrable to chaotic transitions, have been studied in Refs.~\cite{Miller99,Fujisaki03,Bandyopadhyay04,Gammal07,Trail08}. Typically the entanglement is enhanced if the initial unentangled states are localized in chaotic rather than regular phase space regions.  A cold atom experiment may well have observed this sensitivity of entanglement to quantum chaos~\cite{Choudhury09}.

In contrast, this paper addresses the entanglement and localization eigenstate properties of two strongly chaotic subsystems as a function of their interaction strength.  One example could be two particles in a chaotic quantum dot, e.g.~shaped like the Bunimovich stadium~\cite{Bohigas84}.  Or the system might have no obvious classical limit, but possess random-matrix-like quantum fluctuations, e.g.~nonintegrable spin chains~\cite{Pals94,Kudo04}.  Without interactions and although their eigenstates are very complex and delocalized, they are unentangled.  Coupling the particles generates entanglement.  In the strongly interacting limit,  entanglement becomes nearly the maximal possible~\cite{Lakshminarayan01}, similarly to random states in product spaces~\cite{Page93,Bengtsson}.  Recently, it was found that for increasing interaction strength a transition in spectral fluctuations from Poisson-like to random-matrix-like is universal and governed by a dimensionless transition parameter, $\Lambda$~\cite{Srivastava15}.

We derive a surprisingly simple accurate analytical approximation of the transition from unentangled to nearly maximally entangled eigenstates as a function of the same $\Lambda$.  This manifests itself as a simple exponential form for the von-Neumann and Havrda-Charv\'at-Tsallis entropies.  Chaotic dynamics, present even in the non-interacting limit, makes this possible as it allows the application of a combination of random matrix theory (RMT), and a recursively invoked perturbation theory.  In addition, there is a localization transition measurable with the inverse participation ratio (IPR).  By subsystem averaging, a new and intimate connection to entanglement as quantified by the purity emerges.

{\it Bipartite Systems.}---Two bipartite models are considered, an RMT ensemble recently introduced~\cite{Srivastava15}, and a dynamical system consisting of two coupled kicked rotors~\cite{Froeschle72,Lakshminarayan01,Richter14}. In both, the unitary Floquet (time-evolution) operators are of the form ${\cal U}=(U_1 \otimes U_2)\, U_{12}$, where $U_1$ and $U_2$ are subsystem unitary operators on $N$-dimensional Hilbert spaces and $U_{12}$ the entangling interaction in the tensor product space of $N^2$ dimensions.  The eigenproperties for either class of models follow from ${\cal U}|\phi_j\rangle = e^{i \varphi_j} |\phi_j \rangle$.  Equal subsystem dimensionality is studied, but the generalization is immediate.

In the RMT ensemble, $U_1$ and $U_2$ are independently taken from the $N$-dimensional circular unitary ensemble (CUE) matrices~\cite{MehtaBook}, whereas $U_{12}$ is a diagonal matrix whose nonzero elements are of the form $\exp(2 \pi i \epsilon  \xi_{kl})$, where $\xi_{kl}$ ($1\le k,l\le N$) is uniformly distributed in $(-1/2,1/2]$. Here $0 \le \epsilon \le 1$, and $\epsilon=0$ represents no coupling, whereas $\epsilon=1$ implies maximal coupling. Such RMT operators are denoted as ${\cal U}_{\text{RMT}}(\epsilon)$.

The subsystem Floquet operators of the dynamical system are $U_j=\exp(-i {\hat p}_j^2/(2 \hbar)) \exp(-i {\hat V}_j(q_j)/\hbar)$ ($j=1,2$), and $U_{12}=\exp(-i b \, {\hat V}_{12}(q_1,q_2)/\hbar)$ is the interaction, and $b$ an interaction strength. For the kicked rotors, $V_j=K_j \cos(2 \pi q_j)/4 \pi^2$ and interaction $ V_{12}= \cos[2 \pi(q_1+q_2)]/4 \pi^2$. The interaction is diagonal in the position representation, and motivates the simple form of the RMT model. The classical limit of such operators is a 4-dimensional symplectic map~\cite{Froeschle72}.

The individual uncoupled rotors are strongly chaotic with Lyapunov exponents $\approx \ln (K_j/2)$ for large $\{K_j\}$~\cite{Chirikov79}.  The values $K_1=9$ and $K_2=10$ lead to islands of regularity too tiny to influence the quantum spectra perceptibly.  Quantizing unit area phase space tori gives Hilbert spaces of dimension $N$ for each rotor and the scaled Planck constant $h=1/N$.  Including an interaction as above has been studied in different contexts~\cite{Lakshminarayan01,Richter14} where more details are given.  The quantum boundary conditions are chosen to break both parity and time-reversal symmetries. The Floquet operators are denoted as ${\cal U}_{\text{KR}}(b)$.

{\it Universal transition of entropies.}---The mean square interaction matrix element divided by the mean level density squared, $\Lambda$, was given as~\cite{Srivastava15}
\begin{equation}
\Lambda[{\cal U}_{\text{RMT}}(\epsilon)] = \frac{\epsilon^2 N^2}{12}, \;\; \Lambda[{\cal U}_{\text{KR}}(b)] =  \frac{N^4 b^2}{32 \pi^4}.
\end{equation}
The nearest neighbor spacing is Poissonian for $\Lambda=0$ and transitions to the CUE result for $\Lambda \sim 1$. The transition parameter and universal transitions have been observed previously when fundamental or dynamical symmetries are broken~\cite{Pandey83,French88a,Bohigas93,Bohigas95,Cerruti03,Michler12}. We show that $\Lambda$ also governs the entanglement and localization in the eigenvectors $|\phi_j\rangle$, with $\Lambda=0$ corresponding to unentangled states while for $\Lambda \sim 1$ the states are nearly maximally entangled.

As a full system state is pure, its entanglement is characterized by the reduced density matrix eigenvalues~\cite{NielsenChuang10}.  Denote it for the first subsystem with the eigenstate labeled $j$ as $\rho_j ={\text{tr}}_2 (|\phi_j \rangle \langle \phi_j |)$.  With this notation, the von-Neumann entropy $S_1= -{\text{tr}}_1 (\rho_j \ln \rho_j)$ is considered a unique measure~\cite{Bennett96} as it quantifies the entanglement that can be distilled by local operations. The so-called Havrda-Charv{\' a}t-Tsallis entropies~\cite{Havrda67,Tsallis88,Bengtsson}
\begin{equation}
S_k = \dfrac{1-P_k}{k-1}
\label{eq:Tsallis}
\end{equation}
are related to the $k$-th order moments $P_k={\text{tr}}_1( \rho_j ^k)$.  The purity $P_2$ (corresponding to the linear entropy $S_2$)  is often used as a simpler measure of entanglement than the von-Neumann entropy, which emerges in the $k \rightarrow 1$ limit. The eigenstate $|\phi_j \rangle$ is unentangled iff the reduced density matrix $\rho_j$ is pure, in which case all the $S_k$ vanish.

Ahead it is shown that the transition with $\Lambda$ is captured by the entropies in a remarkably simple form
\begin{equation}
\langle S_k(\Lambda) \rangle=\left[1-\exp\left(- \frac{\alpha(k)}{ \langle S_k^{\infty} \rangle}\sqrt{\Lambda}\right) \right] \langle S^{\infty}_k \rangle,
\label{eq:entropy}
\end{equation}
where
\begin{equation}
\alpha(k)=\pi \frac{\Gamma(k-\frac{1}{2})}{\Gamma(k)},\;\;\langle S_k^{\infty} \rangle =\frac{1-C_k N^{1-k}}{k-1}
\label{eq:alphaSinf},
\end{equation}
which are solutions of the equations
\begin{equation}
\label{diffeq}
\frac{ \partial P_k}{ \partial \sqrt{\Lambda}}=-(k-1) \alpha(k)\, \frac{P_k-P_k^{\infty}}{1-P_k^{\infty}}.
\end{equation}
The $C_k$ are Catalan numbers, the  $\langle \cdot \rangle$ represent an ensemble or spectral average, and  $P_k^{\infty}=C_k N^{1-k}$ are moments of the Marcenko-Pastur distribution that determines the large $N$ density of the eigenvalue of $\rho_j$ in the fully interacting RMT limit~\cite{Sommers04}. The asymptotic entropies $\langle S_k^{\infty} \rangle$ are reached at the end of the transition, and whereas Eq.~(\ref{eq:alphaSinf}) is valid for $k>1$, $\langle S_1^{\infty} \rangle =\ln N-\frac{1}{2}$.

\begin{figure}[b]
\includegraphics[scale=1]{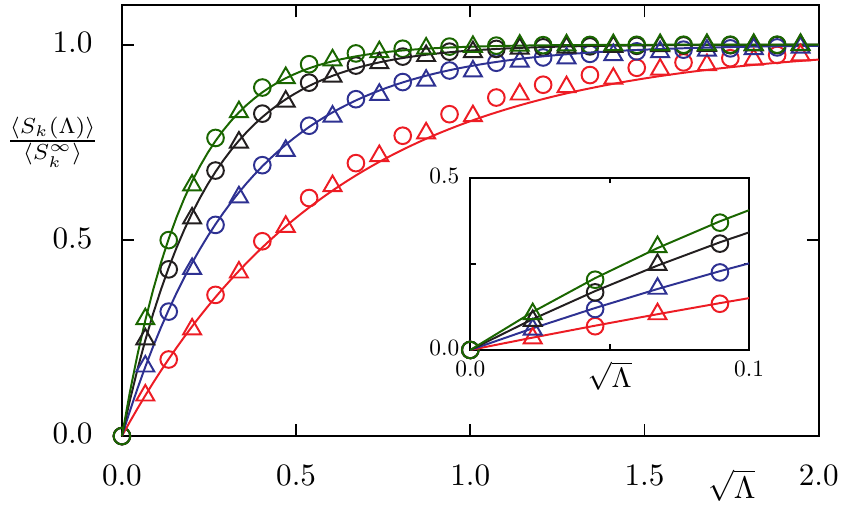}
\caption{\label{fig:entropy} The average eigenstate entropies $\langle S_k \rangle$ for von Neumann, and $k=2,3,4$ as a function of $\sqrt{\Lambda}$. The lowest curve is for von Neumann, and the approach to asymptotic values is faster for larger $k$ values. The triangles are for ${\cal U}_{\text{RMT}}$ and the circles for ${\cal U}_{\text{KR}}$ with $N=50$.  The lines correspond to Eq.~(\ref{eq:entropy}). The inset is a magnification of the small-$\Lambda$ region.}
\end{figure}

Figure~\ref{fig:entropy} shows the von-Neumann entropy along with $S_2,S_3,S_4$.  The agreement with
the RMT ensemble and {\it kicked rotors} is surprisingly good. The entanglement starts from zero and $S_2$ is close to its asymptotic value for $\Lambda \sim 1$.  For high orders, the asymptotic value is roughly reached by $\Lambda \sim k^{-1}$, but for the slowest transition, the von-Neumann entropy, $\Lambda\sim (\log N)^2$.

{\it IPR and purity.}---Localization is a measure of how spread a state $|\psi\rangle$ is in a given basis.  The IPR is defined as $I(\psi )=\sum_{n=1}^{N^2} |\langle n |\psi \rangle |^4$ where $|n\rangle$ is some complete basis.  The maximum of $I(\psi )$ is unity for the most localized state and the minimum is $1/N^2$ for the most delocalized.  For a randomly chosen state of a full bipartite system $I(\psi )$ makes small fluctuations around $2/(N^2+1)$~\cite{UllPor1963}.

\begin{figure}[ht!]
\includegraphics{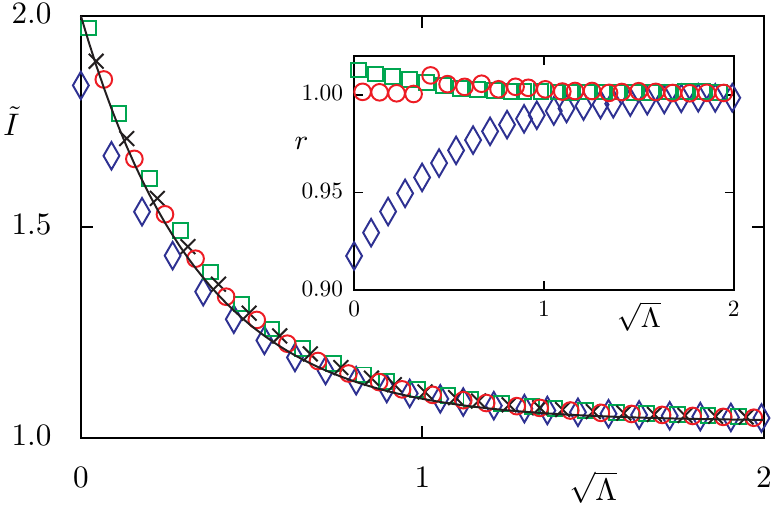}
\caption{\label{fig2} Rescaled spectrally averaged IPR
$\tilde{I} = \tfrac{(N+1)^2}{2} \langle I(\phi_j)\rangle_j$ versus $\sqrt{\Lambda}$
for ${\cal U}_{\text{KR}}(b)$ using position space (blue diamonds)
and momentum space (green squares)
as well as the result of one single ${\cal U}_{\text{RMT}}(\epsilon)$ realization (red circles) for $N=50$.
This is compared with the spectrally averaged purity,
$1 + \langle P_2 \rangle_j$
(black crosses) and the theoretical prediction Eq.~\eqref{eq:IPRpurity} using
the approximation~\eqref{eq:entropy} (solid black line).
The inset shows the ratio $r$ of $\tilde{I}$ and
$1 + \langle P_2 \rangle_j$.}
\end{figure}

Often $I(\phi_j)$ of a particular eigenstate $|\phi_j\rangle$ is evaluated using a product basis such as the kicked rotor's position basis.  This basis dependence can be removed by subsystem averaging.  Using the Haar measure independently on the subsystems gives a direct relation to the purity
\begin{equation}
\langle I (\phi_j) \rangle_{\text{prod}} = \frac{2}{(N+1)^2}\left( 1+P_2 \right) \ ;
\label{eq:IPRpurity}
\end{equation}
the Schmidt decomposition~\cite{NielsenChuang10} is useful for deriving this result.  Thus a subsystem-averaged localization transition between the non-interacting and fully interacting cases is governed by the purity.
Averaging the IPR over all basis states, including entangled ones, gives $\langle I\rangle_{\text{global}}=2/(N^2+1)$~\cite{UllPor1963}.  Indeed averaging of Eq.~\eqref{eq:IPRpurity} with $\phi_j$
sampled with the Haar measure of the full bipartite space of states,
renders the left-hand-side $\langle I\rangle_{\text{global}}$, which implies that $\langle P_2\rangle_{\text{global}}= 2N/(N^2+1)$, consistent with Ref.~\cite{Lubkin78}; this provides an alternative derivation of a random state's average purity.  For non-interacting systems $\langle I  (\phi_j)\rangle_{\text{prod}} = 4/(N+1)^2 \approx 2 \langle I \rangle_{\text{global}}$.  Note that localization and entanglement have been related before, but in apparently disparate ways~\cite{Lakshminarayan03,Giraud07,Viola07}.  Combining the exact relation Eq.~\eqref{eq:IPRpurity} with the approximation Eq.~\eqref{eq:entropy} gives the localization transition shown in Fig.~\ref{fig2} (solid line).

Rather than perform the ensemble average over all product bases in Eq.~\eqref{eq:IPRpurity}, consider a spectral average $\langle I(\phi_j)\rangle_j$ by invoking a property of ergodicity~\cite{Pandey79}:
if the particular product basis used to calculate  $I(\phi_j)$, say the system quantization basis, behaves
like a typical product basis according to the Haar measure, then
Eq.~\eqref{eq:IPRpurity} approximately holds, but with small sample fluctuations.

The rescaled $\langle I(\phi_j)\rangle_j$ versus $\sqrt{\Lambda}$ is shown in
Fig.~\ref{fig2} for three cases: i) a single ${\cal U}_{\text{RMT}}(\epsilon)$ realization,
which follows the theory curve and whose importance is to illustrate the sample fluctuations scale versus $\sqrt{\Lambda}$; ii) ${\cal U}_{\text{KR}}(b)$ using a product position basis; and iii) ${\cal U}_{\text{KR}}(b)$ using a product momentum basis.  To a first approximation, the ${\cal U}_{\text{KR}}(b)$ results closely follow the expected behavior.

It is possible to take a closer look by calculating the ratio of $\langle I(\phi_j)\rangle_j$ by the right side of Eq.~(\ref{eq:IPRpurity}) evaluated using the spectrally averaged purity.  The results are expected to fluctuate about unity.  This is shown in the inset to Fig.~\ref{fig2}.  The momentum basis gives results close to within the range of fluctuations seen in the RMT ratio, whereas the position basis shows deviations from unity on a larger scale.  Thus, the position basis cannot be considered a typical product basis relative to the kicked rotor eigenstates.  Indeed, its special properties can be detected in other measures not shown here, such as intensity densities.  This illustrates the sensitive dependence of deviations from Eq.~(\ref{eq:IPRpurity}) using its spectrally averaged version (without product basis Haar averaging) to non-ergodic features in the quantization basis of a dynamical system.  Deviations could be due to effects such as partial symmetry breaking or strong eigenstate scarring.  Thus, Eq.~(\ref{eq:IPRpurity}) can be exploited as a detector of non-ergodic behaviors.

{\it Reduced density matrix eigenvalues.}---Explicit results for the entropies in Eq.~\eqref{eq:entropy} begin by deriving
perturbative expressions for the eigenvalues of the reduced density matrix.  Let $|\psi^1_k \rangle | \psi^2_l \rangle$ be eigenstates of $U_1 \otimes U_2$ with $k, l \in \{1, 2, \dots, N\}$.  The mean
eigenangle spacings of the subsystems are $2\pi /N$, while that of the full system is $2 \pi /N^2$.
Eigenstates that differ in only one index are separated on average by $2 \pi/N$, thus crucially the nearest levels of any given state differ in both indices. The extent over which this holds is of the order of $N$ levels.  With a weak perturbation $U_\epsilon$,  states will mix with neighboring ones, but due to this property they will to an excellent approximation remain Schmidt decomposed in the unperturbed basis, which for any bipartite pure state is of the form $\sum_{j=1}^N \sqrt{\lambda_j} |\phi_j^1 \rangle |\phi_j ^2 \rangle$~\cite{NielsenChuang10} , where $\lambda_j>0$ ($\sum_j \lambda_j=1$) are the eigenvalues of the reduced density matrices and $|\phi_j^{1,2} \rangle$ are their eigenvectors.

Within the limits alluded to above, the $|\phi_j^{1,2} \rangle$ are the unperturbed states $|\psi^{1,2}_k\rangle $ with a relabelling of indices such that they now order states of the combined system, and the intensities in the unperturbed basis are the eigenvalues of either of the reduced density matrices. Note that the eigenvalues of the reduced density matrix of the first subsystem, say $\rho_j$, will be ordered according to $\lambda_1\ge \lambda_2> \cdots$, and before the perturbation is turned on, only $\lambda_1$ is nonzero and equal to $1$. On turning on the perturbation this changes till for fully interacting systems they are statistically distributed according to the Marcenko-Pastur distribution \cite{Sommers04}.

Consider the usual perturbation theory scenario: $H=H_0 +\epsilon\,  V$ for a bipartite system, with $H_0$ a separable Hamiltonian and $V$ providing the interaction. An unentangled eigenstate of $H_0$, say $|\psi^1_k \rangle | \psi^2_l \rangle$, becomes entangled and up to second-order the eigenvalues of $\rho_j$ become
\begin{align}
\lambda_1 &= 1- \epsilon^2 \sum_{k'l' \neq kl} \frac{ |V_{kl,k'l'}|^2}{(E_{kl}-E_{k'l'})^2},\\
\lambda_2 &= \epsilon^2 \frac{ |V_{kl,k''l''}|^2}{(E_{kl}-E_{k''l''})^2}.
\end{align}
With no special selection rules, the matrix elements in the
numerators can be replaced by relevant random variables
as indicated below. It follows that the case when the
energy level at $k^{''}l^{''}$ is  closest in energy to that at $kl$
typically leads to the second largest eigenvalue $\lambda_2$.
It is sufficient to begin by identifying the largest
two eigenvalues of $\rho_j$.

As the noninteracting subsystems are themselves chaotic, the complex matrix elements have Gaussian densities of characteristic variance $v^2$. Thus, the $w=|V_{kl,k'l'}|^2/v^2$ can be treated as random variables with densities $e^{-w}$. With $D$ the mean level spacing, the scaled energy differences $s^\prime=(E_{kl}-E_{k'l'})/D$ behave as a Poissonian spectrum of unit mean spacing.  The dimensionless transition parameter $\Lambda= \epsilon^2 v^2/D^2$ then naturally determines the effective strength of the interaction.

Consequently, the largest eigenvalue's average can be evaluated from
\begin{equation}
\langle \lambda_1 \rangle = 1-  \int_0^{\infty} \int_0^{\infty}\left(1-\frac{s'}{\sqrt{s'^2 +4 \Lambda w}}\right) R_2(s') e^{-w} ds' dw.
\end{equation}
Here a regularization of $2 \Lambda w/s'^2$ to $(1-s'/\sqrt{s'^2+4 \Lambda w})$ is necessary to remove divergences as the spacing goes to zero. It amounts to treating two levels that are coming very close together as a degenerate two-level subspace~\cite{Cerruti03}.  A different quantity, whose perturbation theory leads to a similar expression, has been studied in the context of parametric eigenstate correlators and the fidelity with a broad variety of physical applications~\cite{Bruus96,Alhassid95,Attias95,Kusnezov96,Cerruti03}.  $R_2(s)$ is the sum of all spacing distributions, and is unity for the Poissonian spectrum. The resulting integrals can be done exactly and give
\begin{equation}
\langle \lambda_1 \rangle = 1-\sqrt{\pi \Lambda}.
\end{equation}
Curiously, there are no higher order corrections.

\begin{figure}[h]
\includegraphics[scale=1]{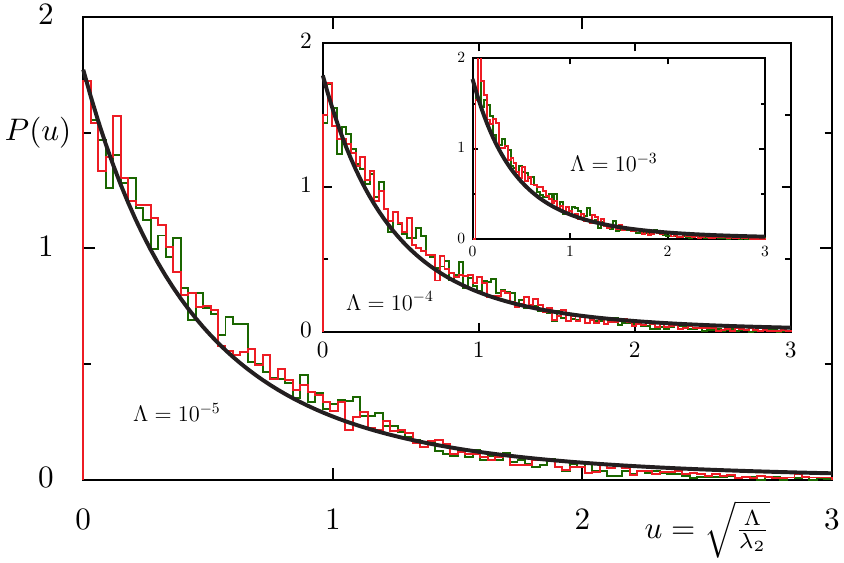}
\caption{\label{fig:distriblam2} The distribution of $u=\sqrt{\Lambda/\lambda_2}$ where $\lambda_2$ is the second largest reduced density matrix eigenvalue of the eigenstates. The solid line is the distribution in Eq.~(\ref{eq:distrlam2}). The data (red histogram) is from the ${\cal U}_{\text{RMT}}$ with $N=50$ and $\Lambda=10^{-5},10^{-4},10^{-3}$ and is (apart from fluctuations)
indistinguishable from the (green) histogram for ${\cal U}_{\text{KR}}$.}
\end{figure}

$\lambda_2$ is a fluctuating random variable of the form $\Lambda w/s^2$, with
$s$ being the nearest neighbor's spacing. In a Poissonian spectrum, the nearest neighbor spacing density $\exp(-s)$ is well-known, but is misleading as for a given level there are two nearest neighbors and it is the smaller which is relevant. With $s=\text{min}(s_1,s_2)$, $P_{\text{min}}(s)=2 \exp(-2s)$ is the required density.
While not required for evaluating $\langle \lambda_2 \rangle$, note the remarkable result that $u =\sqrt{\Lambda/\lambda_2}$ has a universal density independent of $\Lambda$,
\begin{equation}
{\cal P}(u)= 4 \int_0^{\infty} t^2 e^{-t^2} e^{-2 u t} \, dt,
\label{eq:distrlam2}
\end{equation}
for a broad range of $\Lambda$. Of course this is derived within perturbation theory, yet it is excellent for $\Lambda$ over many orders as shown in Fig.~\ref{fig:distriblam2}.

The second largest eigenvalue's average is best evaluated by regularization similar to that of the first,
\begin{equation}
\langle \lambda_2 \rangle= \int_0^{\infty} \int_0^{\infty}\left(1-\frac{s}{\sqrt{s^2 +4 \Lambda w}}\right) e^{-2s} e^{-w} ds \,dw.
\end{equation}
Although possible to give the integrals in terms of special functions, it is more useful to show the small $\Lambda$ expansion
\begin{equation}
\langle \lambda_2 \rangle=\sqrt{\pi \Lambda}+2\Lambda (\gamma+\ln (4\Lambda))-8 \sqrt{\pi} \Lambda^{3/2}+{\mathcal O}(\Lambda^2),
\end{equation}
where $\gamma$ is Euler's constant.
To within perturbation theory $\langle \lambda_1 +\lambda_2 \rangle= 1+{\mathcal O}(\Lambda \ln \Lambda)$,
and it is justifiable to lowest order to consider the changes in the eigenvalues of the reduced density matrix as due to resonances between two nearest neighbors.

{\it Purity and entanglement.}--- With the perturbative expansion for the reduced density matrix eigenvalues the results for the entropies Eq.~\eqref{eq:entropy} and the differential equation Eq.~\eqref{diffeq} can be derived.  A perturbed
state where at most only two of its reduced density matrix eigenvalues are dominant is given by $\sqrt{\lambda_1}|\psi^1_k \rangle | \psi^2_l \rangle+\sqrt{\lambda_2}|\psi^1_{k''} \rangle | \psi^2_{l''} \rangle$. The average purity $ \langle P_2 \rangle =\langle \text{{tr}}(\rho_1^2)\rangle= \langle \lambda_1^2 +\lambda_2^2 \rangle$ can be calculated based on the regularization method used above. With only two levels $\lambda_{1,2}= (1 \pm 1/\sqrt{1+x})/2$ where $x=4\Lambda w/s^2$, leading to
\begin{equation}
\langle P_2 \rangle= 1-\frac{1}{2} \int_0^{\infty}\int_0^{\infty}
\frac{4 \Lambda w}{4 \Lambda w +s'^2} e^{-w} 2 e^{-2s'} \, dw\, ds'.
\end{equation}
The (truly) nearest neighbor spacing $P_{\text{min}}(s)$ is used as before, which reflects the presence of other levels apart from these two. Again the integrals can be evaluated in terms of special functions, but the expansion suffices,
\begin{equation}
\langle P_2 \rangle= 1-\frac{\pi^{3/2}}{2} \sqrt{\Lambda} +  {\mathcal O}(\Lambda \ln \Lambda).
\end{equation}
The following are direct consequences: $\langle \lambda_{1,2}^2 \rangle= 1-\sqrt{\pi \Lambda}\,(1 \pm \pi/4) +{\mathcal O}(\Lambda \ln \Lambda).$
A bit more effort leads to the generalization
\begin{equation}
\langle P_k \rangle= \langle \lambda_1^k +\lambda_2^k \rangle=1- (k-1)\alpha(k) \sqrt{\Lambda} +  {\mathcal O}(\Lambda \ln \Lambda),
\end{equation}
where $\alpha(k)$ is given in Eq.~(\ref{eq:alphaSinf}) and grows with increasing $k$. Hence, the entropies are $\langle S_k \rangle \approx \alpha(k) \sqrt{\Lambda}$ for small $\Lambda$.

For larger $\Lambda$, regimes develop with more than 2 dominant eigenvalues of the reduced density matrix.  To account for these through the full transition, a derivation of an approximate differential equation is useful. Effectively, the perturbation theory can be invoked in a recursive way.  The already superposed state can undergo further mixing with an unentangled state that comes energetically close, to produce say  $\sqrt{\lambda_1^{'}}(\sqrt{\lambda_1}|\psi^1_k \rangle | \psi^2_l \rangle+\sqrt{\lambda_2}|\psi^1_{k''} \rangle | \psi^2_{l''} \rangle)+\sqrt{\lambda_2^{'}}|\psi^1_{k'''} \rangle | \psi^2_{l'''} \rangle$, where $\lambda_{1,2}^{'}$ have the same statistical properties as the unprimed quantities. The purity becomes $P'_2=\lambda_1^{'2} \lambda_1^{2}+\lambda_1^{'2}\lambda_2^{2}+\lambda_2^{'2}$, and the change can be written as $P'_2 -P_2=-(1-\lambda_1^{'2}-\lambda_2^{'2})P_2 +\lambda_2^{'2}(1-P_2)$. Replacing the $\lambda_{1,2}^{'2}$ quantities by their averages leads to
$P_2'-P_2=-\alpha(2) \sqrt{\Lambda}P_2+{\mathcal O}(\Lambda)$. The differentially small $\Lambda$ limit gives the  differential equation $\partial P_2/\partial \sqrt{\Lambda}=-\alpha(2) P_2$. Incorporating finite-$N$ corrections in the large-$\Lambda$ (and large-$N$) limit, where $P_k^{\infty}=C_k/N^{k-1}$, leads to the more correct form, Eq.~(\ref{diffeq}).
Its solution is an exponential decay from $P_k=1$ at $\Lambda=0$.
It is most compactly expressed in terms of the  entropies and leads to Eq.~(\ref{eq:entropy}).
While this is based on a recursively applied perturbation theory for deriving a differential equation and known asymptotics, its accuracy is surprisingly good, see Fig.~\ref{fig:entropy}.

{\it Summary and Outlook.}---As an interaction is introduced, strongly chaotic subsystems develop entanglement in a universal and simple exponential manner.  In fact, each of the entire set of Havrda-Charv{\' a}t-Tsallis entropies has this form. A recursively applied perturbation theory leads to a simple differential equation whose solutions predict the full transition towards entanglement as a function of $\Lambda$.  Furthermore, an exact relation is derived between the subsystem averaged IPR and purity that links entanglement and eigenstate localization properties.  Applying ergodicity, its spectrally averaged variant can be used as a sensitive detector of non-ergodic behaviors.    Extensions to many interacting chaotic systems, while not straightforward, would be extremely interesting from the perspectives of dynamical systems, quantum information, and condensed matter theory.  Another naturally interesting problem concerns the situation when each separable system
shows regular and chaotic dynamics, and the results given here represent an important limit to compare with.

\bibliographystyle{cpg_unsrt_title_for_phys_rev}

\bibliography{quantummodify,rmtmodify,classicalchaos,nano}

\begin{thebibliography}{10}
\newcommand{\enquote}[1]{``#1''}
\providecommand{\url}[1]{\texttt{#1}}
\providecommand{\urlprefix}{URL }
\providecommand{\eprint}[2][]{\url{#2}}

\bibitem{Bennett93}
C.~H. Bennett, G.~Brassard, C.~Cr{\'e}peau, R.~Jozsa, A.~Peres, and W.~K.
  Wootters, \emph{Teleporting an unknown quantum state via dual classical and
  {E}instein-{P}odolsky-{R}osen channels}, Phys.~Rev.~Lett. \textbf{70}, 1895
  (1993).

\bibitem{Amico08}
L.~Amico, R.~Fazio, A.~Osterloh, and V.~Vedral, \emph{Entanglement in many-body
  systems}, Rev.~Mod.~Phys. \textbf{80}, 517 (2008).

\bibitem{Zurek91}
W.~H. Zurek, \emph{Decoherence and the transition from quantum to classical},
  Physics Today \textbf{44}, 36 (1991), also arXiv:quant-ph/0306072.

\bibitem{Zurek03}
W.~H. Zurek, \emph{Decoherence, einselection, and the quantum origins of the
  classical}, Rev.~Mod.~Phys. \textbf{45}, 715 (2003).

\bibitem{Bennett96}
C.~H. Bennett, H.~J. Bernstein, S.~Popescu, and B.~Schumacher,
  \emph{Concentrating partial entanglement by local operations}, Phys.~Rev.~A
  \textbf{53}, 2046 (1996).

\bibitem{Havrda67}
M.~Havrda and F.~Charv{\'a}t, \emph{Quantification method of classification
  processes: {C}oncept of structural $\alpha$-entropy}, Kybernetica \textbf{3},
  30 (1967).

\bibitem{Tsallis88}
C.~Tsallis, \emph{Possible generalization of {B}oltzmann-{G}ibbs statistics},
  J.~Stat.~Phys. \textbf{52}, 479 (1988).

\bibitem{Bengtsson}
I.~Bengtsson and K.~{\.Z}yczkowski, \emph{Geometry of Quantum States: An
  Introduction to Quantum Entanglement}, Cambridge University Press, Cambridge
  (2006).

\bibitem{NielsenChuang10}
M.~A. Nielsen and I.~L. Chuang, \emph{Quantum Computation and Quantum
  Information}, Cambridge University Press, Cambridge (2010).

\bibitem{Albrecht92}
A.~Albrecht, \emph{Investigating decoherence in a simple system}, Phys. Rev. D
  \textbf{46}, 5504 (1992).

\bibitem{Miller99}
P.~A. Miller and S.~Sarkar, \emph{Signatures of chaos in the entanglement of
  two coupled quantum kicked tops}, Phys.~Rev.~E \textbf{60}, 1542 (1999).

\bibitem{Fujisaki03}
H.~Fujisaki, T.~Miyadera, and A.~Tanaka, \emph{Dynamical aspects of quantum
  entanglement for weakly coupled kicked tops}, Phys.~Rev.~E \textbf{67},
  066201 (2003).

\bibitem{Bandyopadhyay04}
J.~N. Bandyopadhyay and A.~Lakshminarayan, \emph{Entanglement production in
  coupled chaotic systems: Case of the kicked tops}, Phys.~Rev.~E \textbf{69},
  016201 (2004).

\bibitem{Gammal07}
A.~Gammal and A.~K. Pattanayak, \emph{Quantum entropy dynamics for chaotic
  systems beyond the classical limit}, Phys.~Rev.~E \textbf{75}, 036221 (2007).

\bibitem{Trail08}
C.~M. Trail, V.~Madhok, and I.~H. Deutsch, \emph{Entanglement and the
  generation of random states in the quantum chaotic dynamics of kicked coupled
  tops}, Phys. Rev. E \textbf{78}, 046211 (2008).

\bibitem{Choudhury09}
S.~Chaudhury, A.~Smith, B.~E. Anderson, S.~Ghose, and P.~S. Jessen,
  \emph{Quantum signatures of chaos in a kicked top}, Nature \textbf{461}, 768
  (2009).

\bibitem{Bohigas84}
O.~Bohigas, M.~J. Giannoni, and C.~Schmit, \emph{Characterization of chaotic
  quantum spectra and universality of level fluctuation laws}, Phys.~Rev.~Lett.
  \textbf{52}, 1 (1984).

\bibitem{Pals94}
P.~van Ede van~der Pals and P.~Gaspard, \emph{Two-dimensional quantum spin
  {H}amiltonians: {S}pectral properties}, Phys.~Rev.~E \textbf{49}, 79 (1994).

\bibitem{Kudo04}
K.~Kudo and T.~Deguchi, \emph{Level statistics of {$XXZ$} spin chains with a
  random magnetic field}, Phys.~Rev.~B \textbf{69}, 132404 (2004).

\bibitem{Lakshminarayan01}
A.~Lakshminarayan, \emph{Entangling power of quantized chaotic systems},
  Phys.~Rev.~E \textbf{64}, 036207 (2001).

\bibitem{Page93}
D.~Page, \emph{Average entropy of a subsystem}, Phys.~Rev.~Lett. \textbf{71},
  1291 (1993).

\bibitem{Srivastava15}
S.~C.~L. Srivastava, S.~Tomsovic, A.~Lakshminarayan, R.~Ketzmerick, and
  A.~B\"acker, \emph{Universal scaling of spectral fluctuation transitions for
  interacting chaotic systems}, arXiv:1509.02329 [nlin.CD]  (2015).

\bibitem{Froeschle72}
C.~Froeschl\'e, \emph{Numerical study of a four-dimensional mapping},
  Astron.~\& Astrophys. \textbf{16}, 172 (1972).

\bibitem{Richter14}
M.~Richter, S.~Lange, A.~B\"acker, and R.~Ketzmerick, \emph{Visualization and
  comparison of classical structures and quantum states of four-dimensional
  maps}, Phys.~Rev.~E \textbf{89}, 022902 (2014).

\bibitem{MehtaBook}
M.~L. Mehta, \emph{Random Matrices (Second Edition)}, Academic Press, London
  (1991).

\bibitem{Chirikov79}
B.~V. Chirikov, \emph{A universal instability of many-dimensional oscillator
  systems}, Phys.~Rep. \textbf{52}, 263 (1979).

\bibitem{Pandey83}
A.~Pandey and M.~L. Mehta, \emph{Gaussian ensembles of random {H}ermitian
  matrices intermediate between orthogonal and unitary ones}, Comm.~Math.~Phys.
  \textbf{87}, 449 (1983).

\bibitem{French88a}
J.~B. French, V.~K.~B. Kota, A.~Pandey, and S.~Tomsovic, \emph{Statistical
  properties of many particle spectra {V}. {F}luctuations and symmetries},
  Ann.~Phys. \textbf{181}, 198 (1988).

\bibitem{Bohigas93}
O.~Bohigas, S.~Tomsovic, and D.~Ullmo, \emph{Manifestations of classical phase
  space structures in quantum mechanics}, Phys.~Rep. \textbf{223}, 43 (1993).

\bibitem{Bohigas95}
O.~Bohigas, M.-J. Giannoni, A.~M. {Ozorio de Almeida}, and C.~Schmit,
  \emph{Chaotic dynamics and the {GOE-GUE} transition}, Nonlinearity
  \textbf{8}, 203 (1995).

\bibitem{Cerruti03}
N.~R. Cerruti and S.~Tomsovic, \emph{A uniform approximation for the fidelity
  in chaotic systems}, J.~Phys. A \textbf{36}, 3451 (2003), corrigenda,
  J.~Phys. A {\bf 36} 11915 (2003).

\bibitem{Michler12}
M.~Michler, A.~B\"acker, R.~Ketzmerick, H.-J. St\"ockmann, and S.~Tomsovic,
  \emph{Universal quantum localizing transition of a partial barrier in a
  chaotic sea}, Phys.~Rev.~Lett. \textbf{109}, 234101 (2012).

\bibitem{Sommers04}
H.-J. Sommers and K.~{\.Z}yczkowski, \emph{Statistical properties of random
  density matrices}, J.~Phys. A \textbf{37}, 8457 (2004).

\bibitem{UllPor1963}
N.~Ullah and C.~E. Porter, \emph{Expectation value fluctuations in the unitary
  ensemble}, Phys.~Rev. \textbf{132}, 948 (1963).

\bibitem{Lubkin78}
E.~Lubkin, \emph{Entropy of an $n$-system from its correlation with a
  $k$-reservoir}, J.~Math.~Phys. \textbf{19}, 1028 (1978).

\bibitem{Lakshminarayan03}
A.~Lakshminarayan and V.~Subrahmanyam, \emph{Entanglement sharing in
  one-particle states}, Phys.~Rev.~A \textbf{67}, 052304 (2003).

\bibitem{Giraud07}
O.~Giraud, J.~Martin, and B.~Georgeot, \emph{Entanglement of localized states},
  Phys.~Rev.~A \textbf{76}, 042333 (2007).

\bibitem{Viola07}
L.~Viola and W.~G. Brown, \emph{Generalized entanglement as a framework for
  complex quantum systems: Purity vs delocalization measures}, J.~Phys.~A
  \textbf{40}, 8109 (2007).

\bibitem{Pandey79}
A.~Pandey, \emph{Statistical properties of many-particle spectra: {III}.
  {Ergodic} behavior in random-matrix ensembles}, Ann.~Phys. \textbf{119}, 170
  (1979).

\bibitem{Bruus96}
H.~Bruus, C.~H. Lewenkopf, and E.~R. Mucciolo, \emph{Parametric conductance
  correlation for irregularly shaped quantum dots}, Phys.~Rev.~B \textbf{53},
  9968 (1996).

\bibitem{Alhassid95}
Y.~Alhassid and H.~Attias, \emph{Universal parametric correlations of
  eigenfunctions in chaotic and disordered systems}, Phys.~Rev.~Lett.
  \textbf{74}, 4635 (1995).

\bibitem{Attias95}
H.~Attias and Y.~Alhassid, \emph{Gaussian random-matrix process and universal
  parametric correlations in complex systems}, Phys.~Rev.~E \textbf{52}, 4776
  (1995).

\bibitem{Kusnezov96}
D.~Kusnezov and D.~Mitchell, \emph{Universal predictions for statistical
  nuclear correlations}, Phys.~Rev.~C \textbf{54}, 147 (1996).

\end{thebibliography}

\end{document}